%% file: Lattice16_bbud_spin.tex
\documentclass{PoS}

\usepackage{dsfont}
\usepackage{epsfig}
\usepackage{graphics}
\usepackage[utf8]{inputenc}

\usepackage{amstext}  
\usepackage{latexsym}  
\usepackage{epsfig}
\usepackage{amssymb}
\usepackage{color}
\usepackage{multirow}
\usepackage{dsfont}

\title{Including heavy spin effects in a lattice QCD study of static-static-light-light tetraquarks}

\ShortTitle{Including heavy spin effects in a lattice QCD study of static-static-light-light tetraquarks}

\author{\speaker{Pedro Bicudo}$^a$, Jonas Scheunert$^b$, Marc Wagner$^b$ \\

$^a$CFTP, Departamento de Física, Instituto Superior Técnico, Universidade de Lisboa, \\
Avenida Rovisco Pais, 1049-001 Lisboa, Portugal \\

$^b$Johann Wolfgang Goethe-Universit\"at Frankfurt am Main, Institut f\"ur Theoretische Physik, \\
Max-von-Laue-Stra{\ss}e 1, D-60438 Frankfurt am Main, Germany \\

E-mail: \email{bicudo@tecnico.ulisboa.pt, scheunert@th.physik.uni-frankfurt.de, mwagner@th.physik.uni-frankfurt.de}
}

\abstract{
In previous works we predicted the existence of a $\bar b \bar b u d$ tetraquark with quantum numbers $I(J^P) = 0(1^+)$ using the static approximation for the $\bar b$ quarks and neglecting heavy spin effects. Since the binding energy is of the same order as expected for these heavy spin effects, it is essential to include them in the computation. Here we present a corresponding method and show evidence that binding is only slightly weakened and that the $\bar b \bar b u d$ tetraquark persists.
}

\FullConference{The 34th International Symposium on Lattice Field Theory, \\
		24-30 July, 2016, \\
		Southampton, UK.}

\begin{document}


\section{Motivation}

Possibly existing heavy-heavy-light-light tetraquarks are currently a ``hot topic'' both experimentally and theoretically, in particular since the observation of the electrically charged $Z_b$ states by the BELLE collaboration in 2011 \cite{Belle:2011aa}.

In previous papers we predicted the existence of a $\bar b \bar b u d$ tetraquark with quantum numbers $I(J^P) = 0(1^+)$ ($I$: isospin; $J$: total angular momentum; $P$: parity) using the static approximation for the $\bar b$ quarks and neglecting heavy spin effects \cite{Bicudo:2012qt,Bicudo:2015vta,Bicudo:2015kna}. Since the obtained binding energy $\Delta E = 90^{+43}_{-36} \, \textrm{MeV}$ is of the same order as expected for heavy spin effects ($\mathcal{O}(m_{B^\ast} - m_B) = \mathcal{O}(46 \, \textrm{MeV})$), it is essential to include heavy spin effects in the computation. In section~\ref{SEC001} we summarize our previous work and in section~\ref{SEC002} we propose a method to take heavy spin effects into account. We also show strong evidence that the $\bar b \bar b u d$ tetraquark persists with only a slightly reduced binding energy $\Delta E = 59^{+30}_{-38} \, \textrm{MeV}$. Parts of this work have been published in \cite{Scheunert:2015pqa}.

Related papers studying also $\bar b \bar b u d$ 4-quark systems are \cite{Stewart:1998hk,Michael:1999nq,Cook:2002am,Doi:2006kx,Detmold:2007wk,Wagner:2010ad,Bali:2010xa,
Wagner:2011ev,Brown:2012tm,Wagenbach:2014oxa,Peters:2015tra,Peters:2016wjm} (static $\bar b$ quarks) and \cite{Francis:2016hui,Peters:2016isf} (NRQCD treatment of $\bar b$ quarks).


\section{\label{SEC001}$\bar b \bar b u d$ tetraquarks, heavy spin effects neglected}

\begin{figure}[htb]
\begin{center}
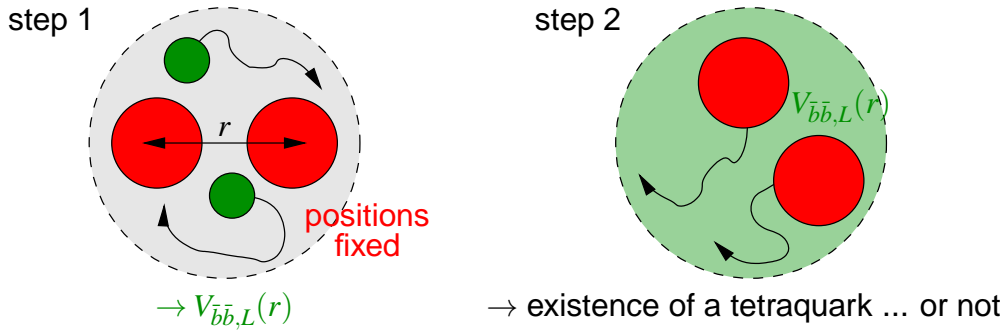
\caption{\label{FIG001}the Born-Oppenheimer approximation for $\bar b \bar b u d$ 4-quark systems.}
\end{center}
\end{figure}

The basic idea of our approach is to investigate the existence of heavy tetraquarks with quark content $\bar b \bar b u d$ in two steps (cf.\ also Figure~\ref{FIG001}):
\begin{itemize}
\item[(1)] Compute potentials of two static antiquarks $\bar b \bar b$ in the presence of two light quarks $u d$ using lattice QCD.

\item[(2)] Check, whether these potentials are sufficiently attractive to host a bound state by solving a corresponding Schr\"odinger equation. A bound state would indicate a stable $\bar b \bar b u d$ tetraquark.
\end{itemize}
This so-called Born-Oppenheimer approximation \cite{Born:1927} is appropriate, if $m_{u,d} \ll m_b$, which is clearly the case for physical quark masses.

\subsection{Born-Oppenheimer approximation, step~(1)}

Step~(1), the lattice QCD computation of $\bar b \bar b$ potentials $V_{\bar b \bar b , L}(r)$, is explained in detail in \cite{Bicudo:2015kna}. It is based on several gauge link ensembles generated by the European Twisted Mass Collaboration (ETMC) with 2 dynamical quark flavors (cf.\ e.g.\ \cite{Boucaud:2008xu,Baron:2009wt}) with light $u/d$ quark mass extrapolations to the physical value.

We use $\bar b \bar b u d$ creation operators
\begin{eqnarray}
\nonumber & & \hspace{-0.7cm} \mathcal{O}_{L,S}(\vec{r}_1,\vec{r}_2) \ \ = \ \ (\mathcal{C} L)_{A B} (\mathcal{C} S)_{C D} \Big(\bar{Q}_C(\vec{r}_1) q_A^{(1)}(\vec{r}_1)\Big) \Big(\bar{Q}_D(\vec{r}_2) q_B^{(2)}(\vec{r}_2)\Big) \quad , \quad r \ \ = \ \ |\vec{r}_1 - \vec{r}_2| , \\
 & &
\end{eqnarray}
where $\bar Q $ denotes a static antiquark representing a $\bar b$ quark. There are quite a number of different channels characterized by
\begin{itemize}
\item isospin, $q^{(1)} q^{(2)} \in \{ (u d - d u) / \sqrt{2} \ \ , \ \ u u , (u d + d u) / \sqrt{2} , d d \}$,

\item light quark spin and parity, $L$ (a $4 \times 4$ matrix acting in spin space),

\item static quark spin and parity (a $4 \times 4$ matrix acting in spin space; irrelevant for $V_{\bar b \bar b , L}(r)$, because static spins are not part of the QCD Hamiltonian).
\end{itemize}
Some of these channels are attractive, others are repulsive, some correspond for large $\bar b \bar b$ separations to pairs of ground state mesons, others correspond to excited mesons (cf.\ Figure~4 and Figure~5 in \cite{Bicudo:2015kna}). As usual in lattice QCD hadron spectroscopy, we compute temporal correlation functions of these creation operators and determine $V_{\bar b \bar b , L}(r)$ for each channel from the exponential decay of its correlation function.

There are two attractive channels corresponding to pairs of ground state mesons, i.e.\ $B$ and/or $B^\ast$ ($B$ and $B^\ast$ are degenerate in the static limit):
\begin{itemize}
\item $I = 0$, $j = 0$ (light spin coupling $L = (\mathds{1} + \gamma_0) \gamma_5$), more attractive;

\item $I = 1$, $j = 1$ (light spin coupling $L = (\mathds{1} + \gamma_0) \gamma_j$), less attractive
\end{itemize}
($j$ denotes the light quark spin). The lattice QCD results can be parameterized by continuous functions using the phenomenologically motivated fitting ansatz
\begin{eqnarray}
\label{EQN003} V_{\bar b \bar b , L}(r) \ \ = \ \ -\frac{\alpha}{r} \exp\bigg(-\bigg(\frac{r}{d}\bigg)^p\bigg)
\end{eqnarray}
with fitting parameters $\alpha$, $d$ and $p$ (one-gluon-exchange at short separations, exponential screening at large separations; cf.\ section~II.B. in \cite{Bicudo:2012qt} for a detailed discussion).


\subsection{Born-Oppenheimer approximation, step~(2)}

In step~(2) we solve the Schr\"odinger equation for the relative coordinate $\vec{r}$ of the two $\bar b$ quarks,
\begin{eqnarray}
\bigg(-\frac{1}{2 \mu} \triangle + V_{\bar b \bar b , L}(r)\bigg) \psi(\vec{r}) \ \ = \ \ E \psi(\vec{r}) \quad , \quad \mu \ \ = \ \ m_b/2 ,
\end{eqnarray}
where $V_{\bar b \bar b , L}(r)$ is one of the two potentials (\ref{EQN003}) obtained in step~(1) and $m_b = 4977$ (from the quark model \cite{Godfrey:1985xj}). Each possibly existing bound state, i.e.\ each eigenvalue $E < 0$, would indicate a stable $\bar b \bar b u d$ tetraquark.

We find only one bound state for one specific potential $V_{\bar b \bar b , L}(r)$, the more attractive potential corresponding to $I = 0$, $j_z = 0$. The binding energy is $\Delta E = -E = 90_{-36}^{+43} \, \textrm{MeV}$, i.e.\ the confidence level for the existence of the $\bar b \bar b u d$ tetraquark is around $2 \, \sigma$. Its quantum numbers are $I(J^P) = 0(1^+)$ as outlined in the following:
\begin{itemize}
\item $\bar b \bar b$ is flavor symmetric, it must be in a color triplet, i.e.\ antisymmetric (otherwise one gluon exchange would not lead to attraction)
\\ $\rightarrow$ due to the Pauli principle the heavy spin must be symmetric, i.e.\ $j_b = 1$.

\item $u d$ with $I = 0$ is flavor antisymmetric, it must be in a color triplet, i.e.\ antisymmetric (otherwise the 4-quark system would not be in a color singlet)
\\ $\rightarrow$ due to Pauli principle the light spin must be antisymmetric, i.e.\ $j = 0$.

\item $j_b = 1$, $j = 0$ and angular momentum $l = 0$ (the bound state corresponds to an $s$ wave) lead to total angular momentum $J = 1$.

\item Ground state mesons $B$ and $B^\ast$ both have $P = -$, the $s$ wave has $P = +$. Therefore, the 4-quark system has $P = +$.
\end{itemize}


\section{\label{SEC002}$\bar b \bar b u d$ tetraquarks, heavy spin effects taken into account}

To take heavy spin effects into account, we first interpret static-static-light-light creation operators and the corresponding potentials $V_L(r)$, $r = |\vec{r}_1 - \vec{r}_2|$ in terms of two heavy-light mesons using
\begin{eqnarray}
\nonumber & & \hspace{-0.7cm} \mathcal{O}_{L,S}(\vec{r}_1,\vec{r}_2) \ \ = \ \ (\mathcal{C} L)_{A B} (\mathcal{C} S)_{C D} \Big(\bar{Q}_C(\vec{r}_1) q_A^{(1)}(\vec{r}_1)\Big) \Big(\bar{Q}_D(\vec{r}_2) q_B^{(2)}(\vec{r}_2)\Big) \ \ = \\
\label{EQN001} & & = \ \ \mathds{G}(S,L)_{a b} \Big(\bar{Q}(\vec{r}_1) \Gamma^{a} q^{(1)}(\vec{r}_1)\Big) \Big(\bar{Q}(\vec{r}_2) \Gamma^{b} q^{(2)}(\vec{r}_2)\Big) ,
\end{eqnarray}
where $\mathds{G}(S,L)_{a b}$ are coefficients, which can be computed using the Fierz identity. Since we use static quarks $\bar Q$ (with only two non-vanishing spinor components), there are 8 possibilties both for $\Gamma^a$ and $\Gamma^b$. The relation to quantum numbers and heavy-light mesons is the following:
\begin{itemize}
\item $\Gamma^{a,b} = (\mathds{1} + \gamma_0) \gamma_5$ $\quad \rightarrow \quad$ $J^P = 0^-$ (the pseudoscalar $B$ meson).

\item $\Gamma^{a,b} = (\mathds{1} + \gamma_0) \gamma_j$ ($j = 1,2,3$) $\quad \rightarrow \quad$ $J^P = 1^-$ (the vector $B^\ast$ meson).

\item $\Gamma^{a,b} = (\mathds{1} + \gamma_0) \mathds{1}$ $\quad \rightarrow \quad$ $J^P = 0^+$ (the scalar $B_0^\ast$ meson).

\item $\Gamma^{a,b} = (\mathds{1} + \gamma_0) \gamma_j \gamma_5$ ($j = 1,2,3$) $\quad \rightarrow \quad$ $J^P = 1^+$ (the pseudovector $B_1^\ast$ meson).
\end{itemize}

In this work we focus on $B$ and $B^\ast$ mesons (the two lightest heavy-light mesons), which are degenerate in the static limit and have similar mass in nature ($m_{B^\ast} - m_B \approx 45 \, \textrm{MeV}$). One can show that there are 16 posibilities of light and static spin couplings,
\begin{eqnarray}
L \, , \, S \in \{ (\mathds{1} + \gamma_0) \gamma_5 \, , \, (\mathds{1} + \gamma_0) \gamma_j \} ,
\end{eqnarray}
which generate exclusively $B$ and/or $B^\ast$ mesons. The corresponding potentials depend only on $L$,
\begin{itemize}
\item $V_5(r) \equiv V_{\bar b \bar b , (\mathds{1} + \gamma_0) \gamma_5}(r)$, i.e.\ $L = (\mathds{1} + \gamma_0) \gamma_5$,
\\ is attractive for isospin $I = 0$, repulsive for isospin $I = 1$,

\item $V_j(r) \equiv V_{\bar b \bar b , (\mathds{1} + \gamma_0) \gamma_j}(r)$, i.e.\ $L = (\mathds{1} + \gamma_0) \gamma_j$,
\\ is repulsive for isospin $I = 0$, attractive for isospin $I = 1$.
\end{itemize}
Neither for $V_5(r)$ nor for $V_j(r)$ it is possible to choose $S$ such that it corresponds exclusively to a $B$ meson pair. One always finds linear combinations of $B$ and $B^\ast$ mesons, e.g.\ for $L = S = (\mathds{1} + \gamma_0) \gamma_5$
\begin{eqnarray}
B(\vec{r}_1) B(\vec{r}_2) + B_x^\ast(\vec{r}_1) B_x^\ast(\vec{r}_2) + B_y^\ast(\vec{r}_1) B_y^\ast(\vec{r}_2) + B_z^\ast(\vec{r}_1) B_z^\ast(\vec{r}_2)
\end{eqnarray}
(the indices $x,y,z$ denote the spin orientation of the $B^\ast$ meson). Vice versa, a $B(\vec{r}_1) B(\vec{r}_2)$ pair does not have defined light quark spin and hence corresponds to a mixture of both $V_5(r)$ or $V_j(r)$, i.e.\ a mixture of an attractive and a repulsive potential.

To study this interplay between the mass difference of $B$ and $B^\ast$ on the one hand and the attractive and repulsive potentials $V_5(r)$ and $V_j(r)$ on the other hand, we consider a coupled channel Schr\"odinger equation for the two $\bar b$ quarks,
\begin{eqnarray}
\label{EQN002} H \Psi(\vec{r}_1,\vec{r}_2) \ \ = \ \ \Big(H_0 + H_{\scriptsize \textrm{int}}\Big) \Psi(\vec{r}_1,\vec{r}_2) \ \ = \ \ E \Psi(\vec{r}_1,\vec{r}_2)
\end{eqnarray}
with a 16-component wave function $\Psi \equiv (
B(\vec{r}_1) B(\vec{r}_2) \ , \
B(\vec{r}_1) B_x^\ast(\vec{r}_2) \ , \
\ldots \ , \
B_z^\ast(\vec{r}_1) B_z^\ast(\vec{r}_2)
)$. The free part of the Hamiltonian is
\begin{eqnarray}
H_0 \ \ = \ \ \frac{\vec{p}_1^2}{2 m_b} \mathds{1}_{16 \times 16} + \frac{\vec{p}_2^2}{2 m_b} \mathds{1}_{16 \times 16} + M \otimes \mathds{1}_{4 \times 4} + \mathds{1}_{4 \times 4} \otimes M
\end{eqnarray}
with $M = \textrm{diag}(m_B \, , \, m_{B^\ast} \, , \, m_{B^\ast} \, , \, m_{B^\ast})$, $m_b = 4977$ (from the quark model \cite{Godfrey:1985xj}) and $m_B = 5280 \, \textrm{MeV}$, $m_{B^\ast} = 5325 \, \textrm{MeV}$ from the PDG \cite{PDG}. The interacting part of the Hamiltonian is
\begin{eqnarray}
H_{\scriptsize \textrm{int}} \ \ = \ \ T^{-1} V(r) T \quad , \quad V(r) \ \ = \ \ \textrm{diag}\Big(\underbrace{V_5(r),\ldots,V_5(r)}_{4 \times} , \underbrace{V_j(r),\ldots,V_j(r)}_{12 \times}\Big) ,
\end{eqnarray}
where $T$ is the transformation between the 16 components of $\Psi$ and the 16 static-static-light-light channels defined by $S$ and $L$ ($T$ is equivalent to the coefficients $\mathds{G}(S,L)_{a b}$ in eq.\ (\ref{EQN001})).

Due to rotational symmetry the coupled channel Schr\"odinger equation (\ref{EQN002}) can be transformed to block diagonal structure, i.e.\ the $16 \times 16$ problem separates into
\begin{itemize}
\item a $2 \times 2$ problem (corresponding to $J = 0$),

\item a $2 \times 2$ and a $1 \times 1$ problem (corresponding to $J = 1$; $3 \times$ degenerate),

\item a $1 \times 1$ problem (corresponding to $J = 2$; $5 \times$ degenerate).
\end{itemize}
Since heavy spin effects will weaken the binding of the $\bar b \bar b u d$ system, it is sufficient to study $I(J^P) = 0(1^+)$, the only channel, for which a $\bar b \bar b u d$ tetraquark has been predicted without taking heavy spin effects into account (cf.\ section~\ref{SEC001}). Since the corresponding $1 \times 1$ problem contains only the repulsive potential $V_j(r)$, it will not have a bound state and can be excluded. The $2 \times 2$ problem, however, contains both the attractive potential $V_5(r)$ and the repulsive potential $V_j(r)$,
\begin{eqnarray}
H_{\scriptsize \textrm{int}, J = 1, 2 \times 2} \ \ = \ \ \frac{1}{2} \left(\begin{array}{cc}
V_5(r) + V_j(r) & V_j(r) - V_5(r) \\
V_j(r) - V_5(r) & V_5(r) + V_j(r)
\end{array}\right) ,
\end{eqnarray}
where the first component of the corresponding wave function $\Psi$ corresponds to $B B^\ast$ and the second component to $B^\ast B^\ast$.

\begin{figure}[htb]
\begin{center}
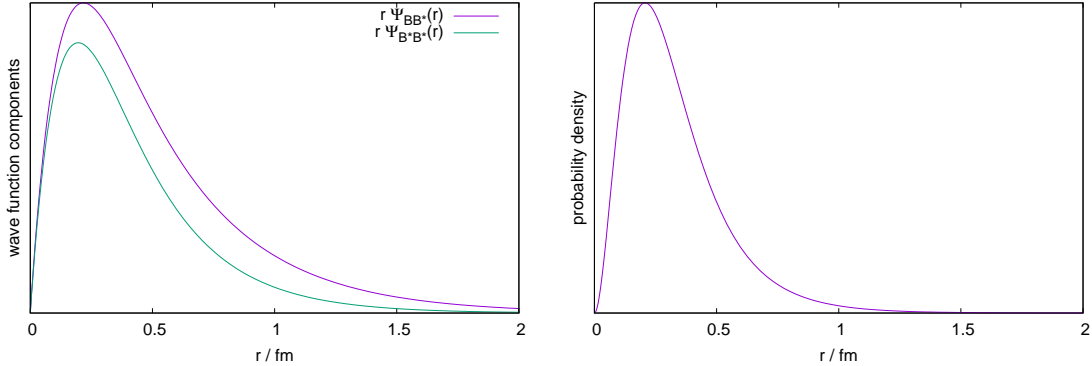
\caption{\label{FIG004}\textbf{(left)}~wave function components of $\Psi$; \textbf{(right)}~probability density for the separation of the two $\bar b$ quarks.}
\end{center}
\end{figure}

This $2 \times 2$ coupled channel Schr\"odinger equation can be solved numerically using a standard Runge-Kutta shooting method. We find that the $\bar b \bar b u d$ tetraquark predicted in section~\ref{SEC001} persists with a slightly reduced binding energy $\Delta E = m_B + m_B^\ast - E = 59^{+30}_{-38} \, \textrm{MeV}$ (without taking heavy spin effects into account $\Delta E = 90^{+43}_{-36} \, \textrm{MeV}$). Consequently the mass of the $\bar b \bar b u d$ tetraquark is $m = m_B + m_B^\ast - \Delta E = (5280 + 5325 - 59_{-30}^{+38}) \, \textrm{MeV} = 10546_{-30}^{+38} \, \textrm{MeV}$. Regarding the structure of the $\bar b \bar b u d$ tetraquark we obtain the following results:
\begin{itemize}
\item The wave function $\Psi$ is a roughly 50\%/50\% mixture of $B B^\ast$ and $B^\ast B^\ast$, i.e.\ the additional energy from the larger mass of the second $B^\ast$ meson is more than compensated by the attractive potential $V_5(r)$ (cf.\ Figure~\ref{FIG004} (left)).

\item The average separation of the two $\bar b$ quarks is around $0.25 \, \textrm{fm}$ (cf.\ Figure~\ref{FIG004} (right)).
\end{itemize}

\acknowledgments

P.B.\ thanks IFT for hospitality and CFTP, grant FCT UID/FIS/00777/2013, for support. M.W.\ acknowledges support by the Emmy Noether Programme of the DFG (German Research Foundation), grant WA 3000/1-1.

This work was supported in part by the Helmholtz International Center for FAIR within the framework of the LOEWE program launched by the State of Hesse.

Calculations on the LOEWE-CSC high-performance computer of Johann Wolfgang Goethe-University Frankfurt am Main were conducted for this research. We would like to thank HPC-Hessen, funded by the State Ministry of Higher Education, Research and the Arts, for programming advice.



\end{document}

%% file: FIG001.eps_t
\begin{picture}(0,0)%
\includegraphics{FIG001.eps}%
\end{picture}%
\setlength{\unitlength}{4144sp}%
\begingroup\makeatletter\ifx\SetFigFont\undefined%
\gdef\SetFigFont#1#2#3#4#5{%
  \reset@font\fontsize{#1}{#2pt}%
  \fontfamily{#3}\fontseries{#4}\fontshape{#5}%
  \selectfont}%
\fi\endgroup%
\begin{picture}(6752,2027)(900,-3437)
\put(3556,-2806){\makebox(0,0)[b]{\smash{{\SetFigFont{12}{14.4}{\sfdefault}{\mddefault}{\updefault}{\color[rgb]{1,0,0}positions}%
}}}}
\put(3556,-2986){\makebox(0,0)[b]{\smash{{\SetFigFont{12}{14.4}{\sfdefault}{\mddefault}{\updefault}{\color[rgb]{1,0,0}fixed}%
}}}}
\put(2701,-3346){\makebox(0,0)[b]{\smash{{\SetFigFont{12}{14.4}{\sfdefault}{\mddefault}{\updefault}{\color[rgb]{0,.56,0}$\rightarrow V_{\bar{b} \bar{b} , L}(r)$}%
}}}}
\put(1666,-1636){\makebox(0,0)[b]{\smash{{\SetFigFont{12}{14.4}{\sfdefault}{\mddefault}{\updefault}{\color[rgb]{0,0,0}step 1}%
}}}}
\put(2701,-2266){\makebox(0,0)[b]{\smash{{\SetFigFont{12}{14.4}{\sfdefault}{\mddefault}{\updefault}{\color[rgb]{0,0,0}$r$}%
}}}}
\put(5806,-3346){\makebox(0,0)[b]{\smash{{\SetFigFont{12}{14.4}{\sfdefault}{\mddefault}{\updefault}{\color[rgb]{0,0,0}$\rightarrow$ existence of a tetraquark ... or not}%
}}}}
\put(4816,-1636){\makebox(0,0)[b]{\smash{{\SetFigFont{12}{14.4}{\sfdefault}{\mddefault}{\updefault}{\color[rgb]{0,0,0}step 2}%
}}}}
\put(6391,-2131){\makebox(0,0)[b]{\smash{{\SetFigFont{12}{14.4}{\sfdefault}{\mddefault}{\updefault}{\color[rgb]{0,.56,0}$V_{\bar{b} \bar{b} , L}(r)$}%
}}}}
\end{picture}%

%% file: FIG004.eps_t
\begin{picture}(0,0)%
\includegraphics{FIG004.eps}%
\end{picture}%
\setlength{\unitlength}{4144sp}%
\begingroup\makeatletter\ifx\SetFigFont\undefined%
\gdef\SetFigFont#1#2#3#4#5{%
  \reset@font\fontsize{#1}{#2pt}%
  \fontfamily{#3}\fontseries{#4}\fontshape{#5}%
  \selectfont}%
\fi\endgroup%
\begin{picture}(6660,2299)(1,-1460)
\end{picture}%

%% file: Lattice16_bbud_spin.bbl
\begin{thebibliography}{99}

\bibitem{Belle:2011aa} 
  A.~Bondar {\it et al.} [Belle Collaboration],
  Phys.\ Rev.\ Lett.\ {\bf 108}, 122001 (2012)
  [arXiv:1110.2251 [hep-ex]].

\bibitem{Bicudo:2012qt} 
  P.~Bicudo and M.~Wagner,
  Phys.\ Rev.\ D {\bf 87}, 114511 (2013)
  [arXiv:1209.6274 [hep-ph]].

\bibitem{Bicudo:2015vta} 
  P.~Bicudo {\it et al.},
  Phys.\ Rev.\ D {\bf 92}, 014507 (2015)
  [arXiv:1505.00613 [hep-lat]].

\bibitem{Bicudo:2015kna} 
  P.~Bicudo, K.~Cichy, A.~Peters and M.~Wagner,
  Phys.\ Rev.\ D {\bf 93}, 034501 (2016)
  [arXiv:1510.03441 [hep-lat]].

\bibitem{Scheunert:2015pqa} 
  J.~Scheunert, P.~Bicudo, A.~Uenver and M.~Wagner,
  Acta Phys.\ Polon.\ Supp.\ {\bf 8}, no.\ 2, 363 (2015)
  [arXiv:1505.03496 [hep-ph]].

\bibitem{Stewart:1998hk}
  C.~Stewart and R.~Koniuk,
  Phys.\ Rev.\ D {\bf 57}, 5581 (1998)
  [arXiv:hep-lat/9803003].

\bibitem{Michael:1999nq}
  C.~Michael and P.~Pennanen [UKQCD Collaboration],
  Phys.\ Rev.\ D {\bf 60}, 054012 (1999)
  [arXiv:hep-lat/9901007].

\bibitem{Cook:2002am}
  M.~S.~Cook and H.~R.~Fiebig,
  arXiv:hep-lat/0210054.

\bibitem{Doi:2006kx}
  T.~Doi, T.~T.~Takahashi and H.~Suganuma,
  AIP Conf.\ Proc.\ {\bf 842}, 246 (2006)
  [arXiv:hep-lat/0601008].

\bibitem{Detmold:2007wk}
  W.~Detmold, K.~Orginos and M.~J.~Savage,
  Phys.\ Rev.\ D {\bf 76}, 114503 (2007)
  [arXiv:hep-lat/0703009].

\bibitem{Wagner:2010ad}
  M.~Wagner [ETM Collaboration],
  PoS LATTICE {\bf 2010}, 162 (2010)
  [arXiv:1008.1538 [hep-lat]].

\bibitem{Bali:2010xa}
  G.~Bali and M.~Hetzenegger,
  PoS {\bf LATTICE2010}, 142 (2010)
  [arXiv:1011.0571 [hep-lat]].
  
  \bibitem{Wagner:2011ev}
  M.~Wagner [ETM Collaboration],
  Acta Phys.\ Polon.\ Supp.\ {\bf 4}, 747 (2011)
  [arXiv:1103.5147 [hep-lat]].

\bibitem{Brown:2012tm}
  Z.~S.~Brown and K.~Orginos,
  Phys.\ Rev.\ D {\bf 86}, 114506 (2012)
  [arXiv:1210.1953 [hep-lat]].
   
\bibitem{Wagenbach:2014oxa}
  B.~Wagenbach, P.~Bicudo and M.~Wagner,
  J.\ Phys.\ Conf.\ Ser.\ {\bf 599}, 012006 (2015)
  [arXiv:1411.2453 [hep-lat]].

\bibitem{Peters:2015tra} 
  A.~Peters {\it et al.},
  PoS LATTICE {\bf 2015}, 095 (2016)
  [arXiv:1508.00343 [hep-lat]].

\bibitem{Peters:2016wjm} 
  A.~Peters, P.~Bicudo, K.~Cichy and M.~Wagner,
  arXiv:1602.07621 [hep-lat].

\bibitem{Francis:2016hui} 
  A.~Francis, R.~J.~Hudspith, R.~Lewis and K.~Maltman,
  arXiv:1607.05214 [hep-lat].

\bibitem{Peters:2016isf} 
  A.~Peters {\it et al.},
  arXiv:1609.00181 [hep-lat].

\bibitem{Born:1927}
  M.~Born and R.~Oppenheimer,
  Annalen der Physik 389, 457 (1927).

\bibitem{Boucaud:2008xu}
  P.~Boucaud {\it et al.} [ETM Collaboration],
  Comput.\ Phys.\ Commun.\ {\bf 179}, 695 (2008)
  [arXiv:0803.0224 [hep-lat]].

\bibitem{Baron:2009wt}
  R.~Baron {\it et al.} [ETM Collaboration],
  JHEP {\bf 1008}, 097 (2010)
  [arXiv:0911.5061 [hep-lat]].

\bibitem{Godfrey:1985xj}
  S.~Godfrey and N.~Isgur,
  Phys.\ Rev.\ D {\bf 32}, 189 (1985).

\bibitem{PDG}
  K.~A.~Olive {\it et al.} [Particle Data Group],
  Chin.\ Phys.\ C, {\bf 38}, 090001 (2014) and 2015 update. 

\end{thebibliography}
